\documentclass[a4paper,fleqn,usenatbib,useAMS,usedcolumn]{mnras}

\usepackage[T1]{fontenc}
\usepackage{ae,aecompl}

\usepackage{graphicx} 

\usepackage{subcaption}
\usepackage{float}

\usepackage[fleqn]{amsmath} 
\usepackage{amssymb}  
\usepackage{gensymb}
\usepackage{hyperref}
\usepackage[nolist]{acronym}
\usepackage{ulem}
\usepackage{etoolbox}
\usepackage[inline]{enumitem}    
\usepackage{arydshln}

\usepackage[usenames, dvipsnames]{color}
\makeatletter
\patchcmd\@combinedblfloats{\box\@outputbox}{\unvbox\@outputbox}{}{%
  \errmessage{\noexpand\@combinedblfloats could not be patched}%
}%
\makeatother

\newcommand{\Msol}[0]{\ensuremath{M_\odot}}
\newcommand{\dd}[0]{\ensuremath{\mathrm{d}}}

\title[Population Parameter Sensitivity]{Accuracy of inference on the physics of binary evolution from gravitational-wave observations}

\author[J.~W.~Barrett et al.]{\parbox{\textwidth}{
Jim W.~Barrett$^{1}$\thanks{E-mail: compas@star.sr.bham.ac.uk },
Sebastian~M.~Gaebel$^{1}$,
Coenraad~J.~Neijssel$^{1}$,
Alejandro~Vigna-G\'omez$^{1}$,
Simon~Stevenson$^{1,2}$,
Christopher~P.~L.~Berry$^{1}$,
Will~M.~Farr$^{1}$ and
Ilya~Mandel$^{1}$
}
\vspace{0.4cm}\\
\parbox{\textwidth}{$^{1}$ Institute of Gravitational-wave Astronomy and School of Physics and Astronomy, University of Birmingham, Edgbaston, Birmingham B15 2TT, United Kingdom}\\
\parbox{\textwidth}{$^{2}$ OzGrav, Swinburne University of Technology, Hawthorn VIC 3122, Australia}
}
\date{\today}

\pubyear{2018}

\begin{document}
\label{firstpage}
\pagerange{\pageref{firstpage}--\pageref{lastpage}}
\maketitle

\begin{abstract}
The properties of the population of merging binary black holes encode some of the uncertain physics underlying the evolution of massive stars in binaries. The binary black hole merger rate and chirp-mass distribution are being measured by ground-based gravitational-wave detectors. We consider isolated binary evolution, and explore how accurately the physical model can be constrained with such observations by applying the Fisher information matrix to the merging black hole population simulated with the rapid binary-population synthesis code \texttt{COMPAS}.  
We investigate variations in four \texttt{COMPAS} parameters: common-envelope efficiency, kick-velocity dispersion, and mass-loss rates during the luminous blue variable and Wolf--Rayet stellar-evolutionary phases.  We find that $\sim1000$ observations would constrain these model parameters to a fractional accuracy of a few per cent. Given the empirically determined binary black hole merger rate, we can expect gravitational-wave observations alone to place strong constraints on the physics of stellar and binary evolution within a few years. 
Our approach can be extended to use other observational data sets; combining observations at different evolutionary stages will lead to a better understanding of stellar and binary physics.
\end{abstract}




\section{Introduction}
\label{sec:Introduction}

Gravitational waves from binary black hole coalescences \citep{BBH:O1,LIGO:GW170608,abbottGW170104,abbottGW170814} have recently been observed by the ground-based gravitational-wave detectors of the Advanced Laser Interferometer Gravitational-Wave Observatory \citep[aLIGO;][]{2015CQGra..32g4001L} and Advanced Virgo \citep[AdV;][]{2015CQGra..32b4001A}. These observations provide a revolutionary insight into the properties of the population of binary black holes. The catalogue of detections will grow rapidly as the instruments continue to improve their sensitivity \citep{abbottObsScen}. In this paper, we analyse how such a catalogue will make it possible to infer the physics of binary evolution by performing inference on parametrised population synthesis models.

A number of channels for the formation of binary black holes have been proposed \citep[see, e.g.,][for reviews]{TheLIGOScientific:2016htt,Miller:2016,MandelFarmer:2017}. In this study, we assume that all merging binary black holes form through classical isolated binary evolution via a common-envelope phase \citep{Postnov2014,Belczynski:2016obo}. While all events observed to date are consistent with having formed through this channel \citep{stevenson2017formation,2017PASA...34...58E,2017arXiv171103556G}, a future analysis would need to hierarchically include the possibility of contributions from multiple channels \citep[e.g.,][]{2017MNRAS.471.2801S,2017ApJ...846...82Z,2017PhRvD..96b3012T}.

Previous efforts to explore how stellar and binary population synthesis models could be constrained with gravitational-wave observations \citep[e.g.,][]{Bulik:2003kr,BulikBelczynski:2003,Mandel:2009nx,Gerosa:2014kta,Stevenson:2015bqa} 
have typically focused on a discrete set of models, usually obtained by varying one evolutionary parameter at a time \citep[e.g.,][]{Voss:2003ep,Dominik:2012,Mennekens:2014}. In this paper, we consider the realistic scenario in which the astrophysical model is described by a multi-dimensional set of continuous parameters which may be strongly correlated. We ask how well we could constrain these parameters with a large observational data set.

The main tool we use to tackle this problem is the Fisher (information) matrix. Fundamentally, if we make an observation of a process, and we have a model for that process that depends on some parameters, then the Fisher matrix quantifies how much we can learn about the parameters in our model from the observation we made. We derive an expression for the Fisher matrix for binary-population synthesis. We use this to quantify how much we can learn about the population parameters from observations of binary black holes using the current generation of ground-based gravitational-wave detectors. While we concentrate on gravitational-wave observations here, the method is applicable to other data sets, and the best constraints may come from combining multiple complementary observations.

We use Fisher matrices to demonstrate that it may be possible to precisely measure the population parameters in binary-population synthesis models with $\sim1000$ observations of binary black hole mergers. At the expected rate of gravitational-wave detections \citep{abbottGW170104}, this could be within a few years of the detectors reaching design sensitivity ($\sim2$--$3~\mathrm{yr}$ at design sensitivity for our fiducial model); the observing schedule for gravitational-wave observatories is given in \citet{abbottObsScen}.

We first give an introduction to our binary population synthesis model in section~\ref{sec::binaryPopSynth}, together with a description of the model parameters we wish to infer using gravitational-wave observations. In section~\ref{sec::modelpredictions}, we demonstrate how we transform the raw outputs of our binary population synthesis model by considering observational selection effects and redshift- and metallicity-dependent star formation rates.
In section~\ref{sec::methods} we introduce the statistical tools used in this paper:
\begin{enumerate*}[label=(\roman*), itemjoin={{, }}, itemjoin*={{, and }}]
    \item the likelihood function representing the probability of an observation given our model
      \item a method for including measurement uncertainties in observations
    \item the Fisher matrix, which quantifies the sensitivity of our model to changes in its underlying parameters.
   \end{enumerate*}
The results of applying this methodology to binary population synthesis models are presented and discussed in section~\ref{sec::results}, and we discuss our conclusions in section~\ref{sec::conclusion}.

\section{Population synthesis of massive stellar binaries}
\label{sec::binaryPopSynth}

Many of the details of binary evolution are currently uncertain \citep{Postnov2014,DeMarcoIzzard:2017}.
Population synthesis models efficiently, albeit approximately, simulate the interactions of a large number of binaries in order to capture population wide behaviour and thoroughly explore the space of initial conditions. Uncertainties in the physics underlying isolated binary evolution are captured within population synthesis models through tunable parameters, which we call population parameters. In this paper we focus on four population parameters which have an impact on binary black hole formation.  We use the rapid population synthesis code \texttt{COMPAS}.\footnote{Further details and sample \texttt{COMPAS} simulations are available at \href{http://www.sr.bham.ac.uk/compas/}{www.sr.bham.ac.uk/compas/}.} This uses the stellar evolutionary models of \citet{hurley2000comprehensive}. Final black hole masses are calculated using the delayed model of \citep{fryer2012compact}. With the exception of the variations to the four population parameters we describe in section~\ref{ssec:populationParameters}, we employ the \citet{stevenson2017formation} fiducial model throughout this paper.

\subsection{Population parameters}
\label{ssec:populationParameters}

\subsubsection{Supernova kick velocity} \label{sssec:vKickSigma}

The asymmetric ejection of matter \citep{JankaMueller:1994,BurrowsHayes:1996,Janka:2013hfa} or emission of neutrinos \citep{Woosley:1987,Bisnovatyi-Kogan:1993,Socrates:2005} during a supernova can provide a kick to the stellar remnant. This birth kick is on the order of hundreds of $\mathrm{km\,s}^{-1}$ for neutron stars \citep{2005MNRAS.360..974H}.  The typical strength of supernova kicks imparted to black holes is not well constrained observationally \citep{2014ApJ...790..119W,2016MNRAS.456..578M,2017MNRAS.467..298R}, although they may be reduced relative to neutron star through the accretion of material falling back onto the stellar remnant \citep{fryer2012compact}.

In \texttt{COMPAS}, the strength of supernova kicks is parametrised using the dispersion parameter for a $3$-dimensional Maxwell--Boltzmann distribution $\sigma_\mathrm{kick}$. A kick velocity $v_\mathrm{kick}$ is drawn from the distribution
\begin{equation}
  P(v_\mathrm{kick}) = \sqrt{\frac{2}{\pi}} \; \frac{v_\mathrm{kick}^2}{\sigma_\mathrm{kick}^3} \exp\left(\frac{-v_\mathrm{kick}^2}{2\sigma_\mathrm{kick}^2}\right).
\label{eq:kick_maxwell}
\end{equation}
Alternative parametrisations for the supernova kick have been considered by \citet{BrayEldridge:2016}, who did not find sufficient evidence to prefer them; here, we only consider continuous variations to model parameters, including the kick velocity in the Maxwell--Boltzmann distribution. 

The kick is modified to account for mass fallback, so that the final kick imparted to the black hole is
\begin{equation}\label{eq::kickFallback}
v^*_\mathrm{kick} = (1-f_\mathrm{fb})v_\mathrm{kick},
\end{equation}
where $f_\mathrm{fb}$ is the fraction of matter that falls back on to the black hole, calculated according to the delayed model of \citet{fryer2012compact}. For carbon--oxygen core masses greater than $11 \Msol$, $f_\mathrm{fb}=1$ and so many heavy black holes receive no natal kick in this model \citep{Belczynski:2016obo,stevenson2017formation}. Whilst observations of the proper motions of isolated Galactic pulsars \citep{Hobbs2005Pulsars} suggest a value of $\sigma_\mathrm{kick} = 265~\mathrm{km\,s}^{-1}$, we choose a fiducial $\sigma_\mathrm{kick} = 250~\mathrm{km\,s}^{-1}$  to match \citet{stevenson2017formation}.

\subsubsection{Common-envelope efficiency}\label{sssec:ceAlpha}

When mass transfer is dynamically unstable and initially proceeds on the very short dynamical timescale of the donor, a shared, non co-rotating common envelope is formed around the donor core and the companion \citep{Paczynski:1976}.  
The details of the common-envelope phase are amongst the least well understood across all phases of isolated binary evolution \citep[for a review, see][]{Ivanova:2013CEReview}.

In \texttt{COMPAS}, the classical energy formalism \citep{Webbink:1984} is employed to parametrise uncertainty in the physics of the common envelope. When a binary begins a common-envelope phase, each envelope is bound to its core, with a total binding energy approximated by
\begin{equation}
E_{\mathrm{bind}} = -G \left[\frac{M_1 (M_1 - M_{\mathrm{core},1})}{\lambda_{\mathrm{CE},1} R_1} + \frac{M_2 (M_2 - M_{\mathrm{core},2})}{\lambda_{\mathrm{CE},2} R_2}\right],
\label{eq:binding_energy_ce}
\end{equation}
where $G$ is Newton's constant, $M_{\mathrm{core},(1,2)}$ are the core masses of the two stars, $M_{(1,2)}$ and $R_{(1,2)}$ are the stellar masses and radii, respectively, and $\lambda_{\mathrm{CE}(1,2)}$ are the corresponding stellar-structure parameters introduced by \cite{deKool:1990} and are functions of star's evolutionary state \citep[e.g.,][]{Dewi:2000nq,Kruckow:2016tti}.

The loss of co-rotation between the orbit of the cores and the common envelope leads to energy dissipation which causes the cores to spiral in. Some of this lost orbital energy may be eventually used to eject the common envelope. The efficiency with which this transfer of energy occurs is uncertain, and is characterised by the free parameter $\alpha_{\mathrm{CE}}$. In order to determine the separation after the common-envelope phase, the classical energy formalism compares the binding energy of the envelope to the energy transferred from the orbit $\Delta E_\mathrm{orbit}$ so that
\begin{equation}
 E_\mathrm{bind}= \alpha_{\mathrm{CE}}\Delta E_\mathrm{orbit}  \, .
\label{eq:alpha_ce}
\end{equation}
If the binary has sufficient orbital energy to completely expel the envelope, we consider this a successful common-envelope event. Unsuccessful ejections lead to a merger before a binary black hole system is formed. We follow \citet{stevenson2017formation} in assuming that common-envelope phases initiated by main sequence of Hertzsprung gap donors always lead to mergers \citep[cf.\ the pessimistic model of][]{Dominik:2012}. 

The fiducial choices of the parameters in \texttt{COMPAS} are $\lambda_{\mathrm{CE}} = 0.1$ and $\alpha_{\mathrm{CE}} = 1.0$. We explicitly leave $\lambda_{\mathrm{CE}}$ fixed whilst making small perturbations to $\alpha_{\mathrm{CE}}$; however, this is an issue of labelling, since it is the product of these two free parameters which is ultimately of importance to the common-envelope physics \citep{Dominik:2012}. 

\subsubsection{Mass-loss multipliers}\label{sssec::massLossMultipliers}

Throughout their lives, stars lose mass through stellar winds. The wind mass-loss rate depends strongly on the star's luminosity and is generally highest for high mass, high metallicity stars. The dearth of observations of low metallicity environments means wind mass-loss rates are poorly constrained at low metallicities, and at high masses where stars are intrinsically rare. These are precisely the regimes where the progenitors of gravitational-wave sources are likely to form \citep{Belczynski:2016obo,2016MNRAS.462.3302E,lamberts2016and,stevenson2017formation,2017arXiv171103556G}.

\texttt{COMPAS} employs the approximate wind mass-loss prescriptions detailed in \citet{belczynski2010maximum}. For hot O/B-stars, we employ the wind mass-loss prescription of \citet{Vink:2001}. Our Wolf--Rayet wind mass-loss rates come from \citet{1998A&A...335.1003H}. For other phases the mass-loss prescriptions from \citet{hurley2000comprehensive} are used. Uncertainty in mass-loss rates can have a significant impact on stellar evolution; for example, \citet{2017A&A...603A.118R} find that there is a $\sim50~\mathrm{per\ cent}$ uncertainty in the mapping between initial and final masses when considering different mass-loss prescriptions when modelling solar-metallicity, non-rotating, single stars, with initial masses between $15 \Msol$ and $35 \Msol$. There are particular phases of stellar evolution where the mass-loss rates lack strong constraints by observations. We parametrise the mass-loss rates in two of these phases with tunable population parameters.

During the luminous blue variable (LBV) phase \citep{1994PASP..106.1025H}, extremely massive stars undergo a relatively short episode of rapid mass loss which strongly impact the binary's future evolutionary trajectory \citep[e.g.,][]{Mennekens:2014}; observational constraints on the physics of LBV stars are currently uncertain \citep{2017RSPTA.37560268S}.\footnote{As in \citet{hurley2000comprehensive}, we assume stars are in an LBV-like phase if their luminosity and radius satisfy $L > 6\times 10^5 L_\odot$ and $(R/R_\odot)(L/L_\odot)^{1/2} > 10^5$.} Following \citet{belczynski2010maximum}, we parametrise this rate in terms of a multiplicative factor $f_{\mathrm{LBV}}$ used to modify the basic prescription, so that the rate becomes
\begin{equation}
\dot{M}_{\mathrm{LBV}} = f_{\mathrm{LBV}} \times 10^{-4}\; \Msol\, \mathrm{yr}^{-1};
\label{eq:mdot_lbv}
\end{equation}
our fiducial value for this factor is $f_{\mathrm{LBV}} = 1.5$ \citep{belczynski2010maximum}.

During the Wolf--Rayet phase, stars have lost their hydrogen envelopes and have high but relatively poorly constrained mass-loss rates \citep{2007ARA&A..45..177C}. We use a multiplicative constant $f_{\mathrm{WR}}$ to modify the base rate:
\begin{equation}
\dot{M}_{\mathrm{WR}} = f_{\mathrm{WR}} \left(\frac{L}{L_{\odot}}\right)^{1.5}\left(\frac{Z}{Z_{\odot}}\right)^m \times 10^{-13}\; \Msol\; \mathrm{yr}^{-1},
\label{eq:mdot_wr}
\end{equation}
where $L$ is the stellar luminosity, $Z$ is the metallicity, $Z_{\odot} = 0.02$ is approximately the bulk metallicity of our Sun, and $m=0.86$ is an empirically determined scaling factor \citep{Vink:2005,belczynski2010maximum}. The fiducial choice for this population parameter is $f_{\mathrm{WR}} = 1.0$. We use the same mass-loss prescription for all Wolf-Rayet subtypes \citep{belczynski2010maximum}, as the \citet{hurley2000comprehensive} evolutionary tracks do not distinguish between them. Recent investigations of mass loss for Wolf--Rayet stars of varying composition include \citet{2016MNRAS.459.1505M,2016ApJ...833..133T,2017MNRAS.470.3970Y}.

\section{Model predictions}
\label{sec::modelpredictions}

In this paper we evaluate the impact of the tunable parameters described above on the rate of detections and the measured chirp-mass distribution of binary black holes. The chirp mass $\mathcal{M}$ is a particular combination of the component masses $M_1, M_2$ which is measured well from the gravitational-wave frequency evolution during the binary inspiral \citep{Cutler:1994ys,2016PhRvL.116x1102A},
\begin{equation}
\mathcal{M} = \frac{(M_1 M_2)^{3/5}}{(M_1 + M_2)^{1/5}}.
\label{eq:chirp_mass}
\end{equation}

The chirp mass is just one of the parameters measurable through gravitational waves, other observables such as component masses, spins and the distance to the source can also be inferred \citep{2016PhRvL.116x1102A}. For simplicity, we have chosen to focus on chirp mass since it is the best measured. This is a conservative approach, as we have neglected information about other parameters; however, the methods presented here are easily extendible to include other observables.

In order to represent the distribution of chirp masses produced by the population synthesis model, we chose to bin our systems by chirp mass. Throughout this paper, we use $30$ bins of equal width, ranging from the lowest to the highest chirp masses present in our dataset. The number of bins is determined by the scale length of variability in the chirp-mass distribution and the chirp-mass measurement uncertainty discussed below; the results are insensitive to halving the number of bins.

The raw output of a population synthesis model is a list of the initial conditions and final outcomes of all the binaries simulated. In order to compare this output to astronomical observations, it is necessary to process the data further, in order to account for the history of star formation in the Universe and the observational selection effects. We describe this processing below.

\subsection{Cosmic history}

\begin{figure}
\begin{center}
 \includegraphics[width=0.5\textwidth]{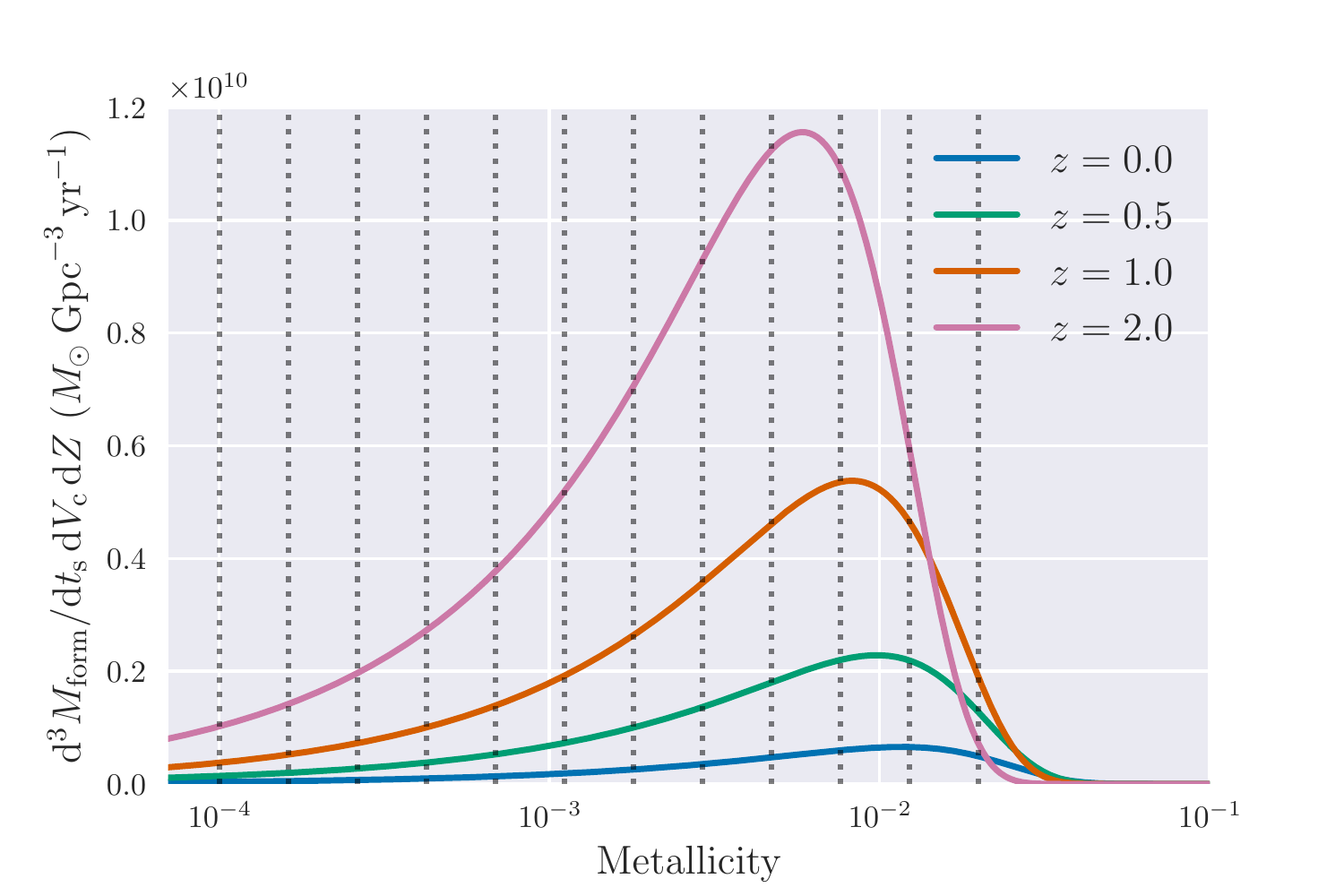}
 \caption{The metallicity-specific star formation rate as a function of metallicity at three different redshifts, using the star-formation-rate distribution of \citet{madau2014cosmic} and the metallicity distribution of \citet{langer2006collapsar}. The vertical dashed lines indicate the metallicities at which we undertook simulations for this study. Metallicities above $Z_\odot=0.02$ contribute negligibly to the binary black hole merger rate.}
   \label{fig::MSSFR}
\end{center}
\end{figure}

\begin{figure}
\begin{center}
 \includegraphics[width=0.5\textwidth]{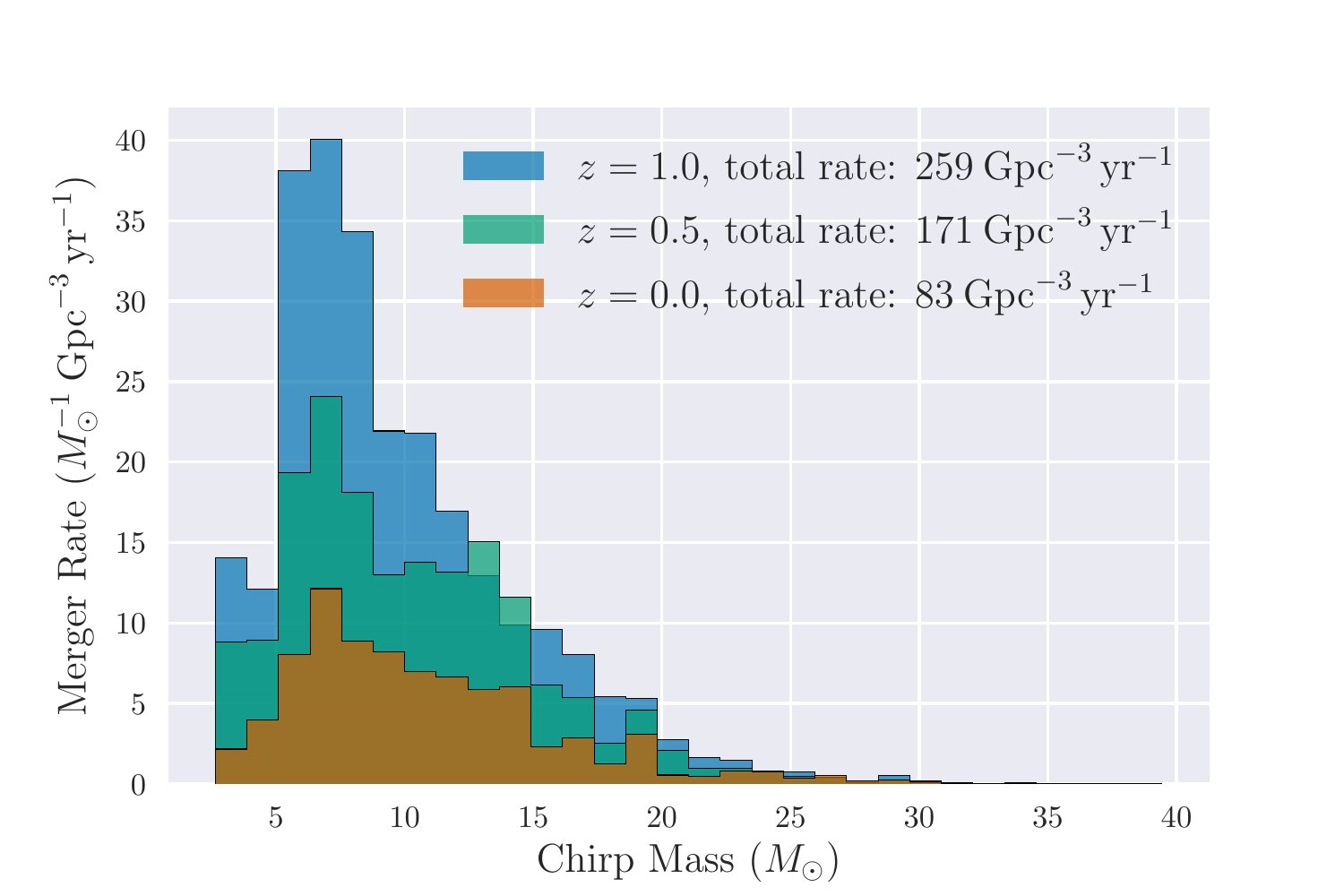}
 \caption{The binary black hole merger rate predicted by the \texttt{COMPAS} fiducial model at three different redshifts, taking into account the cosmic evolution of the metallicity-specific star formation rate. For comparison, the total inferred merger rate density from gravitational-wave observations is $12$--$213~\mathrm{Gpc^{-3}\,yr^{-1}}$ \citep{abbottGW170104}.}
   \label{fig::cosmicRates}
\end{center}
\end{figure}

In order to focus our computation on black hole progenitors, we only simulate systems with primary masses between $7 \Msol<M_1<100 \Msol$.  We assume that all stars are in binaries with primary masses ranging between $0.01$--$150 \Msol$ following the initial mass function of \citet{kroupa2001variation} with a flat mass-ratio distribution \citep{Sana:2012}. At formation, binaries are assumed to have a uniform-in-the-logarithm distribution of orbital separations \citep{1924PTarO..25f...1O,1983ARA&A..21..343A} and zero orbital eccentricity; for more detailed studies of mass-ratio and orbital distributions, see \citet[][]{duchene2013stellar,moe2017mind}.
\texttt{COMPAS} simulations produce a rate of binary black hole formation per unit star formation mass $M_\mathrm{form}$, 
\begin{equation}
\mathcal{R}_\texttt{COMPAS} = \frac{\dd^3 N_{\mathrm{form}}}{\dd M_\mathrm{form}\,\dd\tau_\mathrm{delay}\,\dd\mathcal{M}},
\end{equation}
where $\tau_\mathrm{delay}$ is the delay time, defined as the time from the birth of a binary to its coalescence \citep{Peters:1964}. 
To compute the total rate of binary black hole mergers per unit comoving volume per unit time we need to convolve the \texttt{COMPAS} formation rate with the amount of metallicity-specific star formation per unit volume per unit time at the birth of the binaries. Delay times can range from a few $\mathrm{Myr}$ to $\mathrm{Gyr}$, and observations show that both the metallicity and star formation rates in galaxies evolve significantly over these timescales \citep{madau2014cosmic}. We use the star formation rate distribution of \citet{madau2014cosmic} and the metallicity distribution of \citet{langer2006collapsar}. Other distributions have been suggested \citep[e.g.,][]{savaglio2005gemini,ma2015origin,2015MNRAS.447.2575V}, and the cosmic history of metallicity evolution adds an additional source of uncertainty to our model predictions. Future studies could consider how metallicity evolution could be included with the other model parameters and inferred from binary observations.  In figure~\ref{fig::MSSFR} we provide an illustration of the metallicity-specific star formation rate at redshifts $z=0.5$, 1, and 1.5, and also indicate metallicities at which we performed simulations for this study. We use these to translate the star formation rate into the merger rate at redshift $z$
\begin{align}
\label{eq::cosmicIntegral}
\frac{\dd^3N_{\mathrm{merge}}}{\dd t_{\mathrm{s}} \,\dd V_{\mathrm{c}}\,\dd\mathcal{M}} (z) & =  \int \dd Z \ \int \dd\tau_\mathrm{delay} \left[\frac{\dd^3N_{\mathrm{form}}}{\dd M_{\mathrm{form}}\,\dd\tau_\mathrm{delay}\,\dd\mathcal{M}}(Z) \right. \nonumber\\
& {} \times \left. \frac{\dd^3M_{\mathrm{form}}}{\dd t_{\mathrm{s}} \, \dd V_{\mathrm{c}}\,\dd Z}(Z,t_\mathrm{form}=t_\mathrm{merge}(z) - \tau_\mathrm{delay})\right],
\end{align}
where $t_{\mathrm{s}}$ is the time measured in the frame of reference of the merger, $V_{\mathrm{c}}$ is the comoving volume and we use cosmological parameters from \citet{Planck:2015}. Figure~\ref{fig::cosmicRates} shows the local merger rate at three different redshifts after accounting for changes in star formation rate and cosmology.

\subsection{Selection effects}\label{ssec::selectionEffects}

Gravitational-wave detectors are not equally sensitive to every source. The distance to the source, its orientation and position relative to the detectors, as well as the physical characteristics of the source all affect how likely it is that the system would be detectable. The detectability of a signal depends upon its  signal-to-noise ratio (SNR).
The SNR in a single detector is defined as \citep{1992PhRvD..46.5236F}
\begin{equation}
\mathrm{SNR}^{2} = \left<h\middle|h\right> = 4 \Re \int_{f_\mathrm{min}}^{f_\mathrm{max}} \dd f\; \frac{h^{*}(f)h(f)}{S(f)},
\label{eq:snr}
\end{equation}
where $h(f)$ is the waveform measured by the detector, $S(f)$ is the one-sided noise power spectral density, and $f_\mathrm{min}$ and $f_\mathrm{max}$ are the limits of the frequency range considered. 

For simplicity, we assume that signals are detected if their single-detector SNR exceeds a threshold value of $8$ \citep{abbottObsScen}. To model the waveforms, we use the IMRPhenomPv2 \citep{Hannam:2013oca,Husa:2015iqa,Khan:2015jqa} and SEOBNRv3 \citep{Pan:2013rra,Babak:2016tgq} approximants;\footnote{We use the implementations publicly available in the LAL suite software package \href{https://wiki.ligo.org/DASWG/LALSuite}{wiki.ligo.org/DASWG/LALSuite}.} these include the inspiral, merger and ringdown phases of a binary black hole coalescence, and allow for precession of the black hole spins. We incorporate the effects of cosmological redshift, which manifest as an apparent increase in the system masses, $M_\mathrm{obs} = (1+z)M_\mathrm{s}$ \citep{1987GReGr..19.1163K,2005ApJ...629...15H}. We assume a detector sensitivity equal to aLIGO in its design configuration \citep{2015CQGra..32g4001L,abbottObsScen}.

We optimise our computations, reducing the number of waveform calculations required, by exploiting the fact that the parameters extrinsic to the gravitational-wave source, such as its position and orientation, only couple into the overall amplitude of the wave via
\begin{align}
\mathcal{A} \propto & \, \frac{1}{D_\mathrm{L}}\sqrt{F_+^2(1+\cos^2 i)^2 + 4F_\times^2\cos^2 i},\\
F_+ \equiv & \, \frac{1}{2} \cos(2\psi)[1+\cos^2(\theta)]\cos(2\phi)\nonumber \\
& - \sin(2\psi)\cos(\theta)\sin(2\phi), \\
F_\times \equiv & \, \frac{1}{2} \sin(2\psi)[1+\cos^2(\theta)]\cos(2\phi) \nonumber\\
& + 2\cos(2\psi)\cos(\theta)\sin(2\phi),
\end{align}
where $\mathcal{A}$, $D_\mathrm{L}$, $i$, $\psi$, $\theta$ and $\phi$ are the gravitational-wave amplitude, luminosity distance, inclination, polarization, and polar and azimuthal angles of the source location in the detector frame, respectively \citep{1987GReGr..19.1163K,Cutler:1994ys}. Therefore, we need only compute the phase evolution for a given combination of intrinsic binary parameters, such as masses, once, and then marginalize over the extrinsic parameters (with the exception of $D_\mathrm{L}$) as described in \citet{FinnChernoff1993OneInterferometer}.

For a system with a given $(M_1, M_2, D_\mathrm{L})$, we determine the fraction of  extrinsic parameter realisations for which the observed SNR passes our threshold, and label this as our detection probability $P_\mathrm{det}$.

We can use this detection probability to transform the merge rate given in Eq.~\eqref{eq::cosmicIntegral} into a rate of detections. Integrating over the merger redshift gives the total detection rate
\begin{equation}
\frac{\dd N_{\mathrm{obs}}}{\dd t_{\mathrm{obs}} \,\dd\mathcal{M}} = \; \int \dd z \;\left[ \frac{\dd^3 N_{\mathrm{merge}}}{\dd t_{\mathrm{s}} \,\dd V_{\mathrm{c}}\,\dd\mathcal{M}} \frac{\dd V_{\mathrm{c}}}{\dd z} \frac{\dd t_\mathrm{s}}{\dd t_{\mathrm{obs}}} P_\mathrm{det}\right],
\end{equation}
where $t_\mathrm{s}$ is time in the source frame and $t_{\mathrm{obs}} = (1+z)t_\mathrm{s}$ is time in the observer's frame.

\begin{figure}
\begin{center}
 \includegraphics[width=0.5\textwidth]{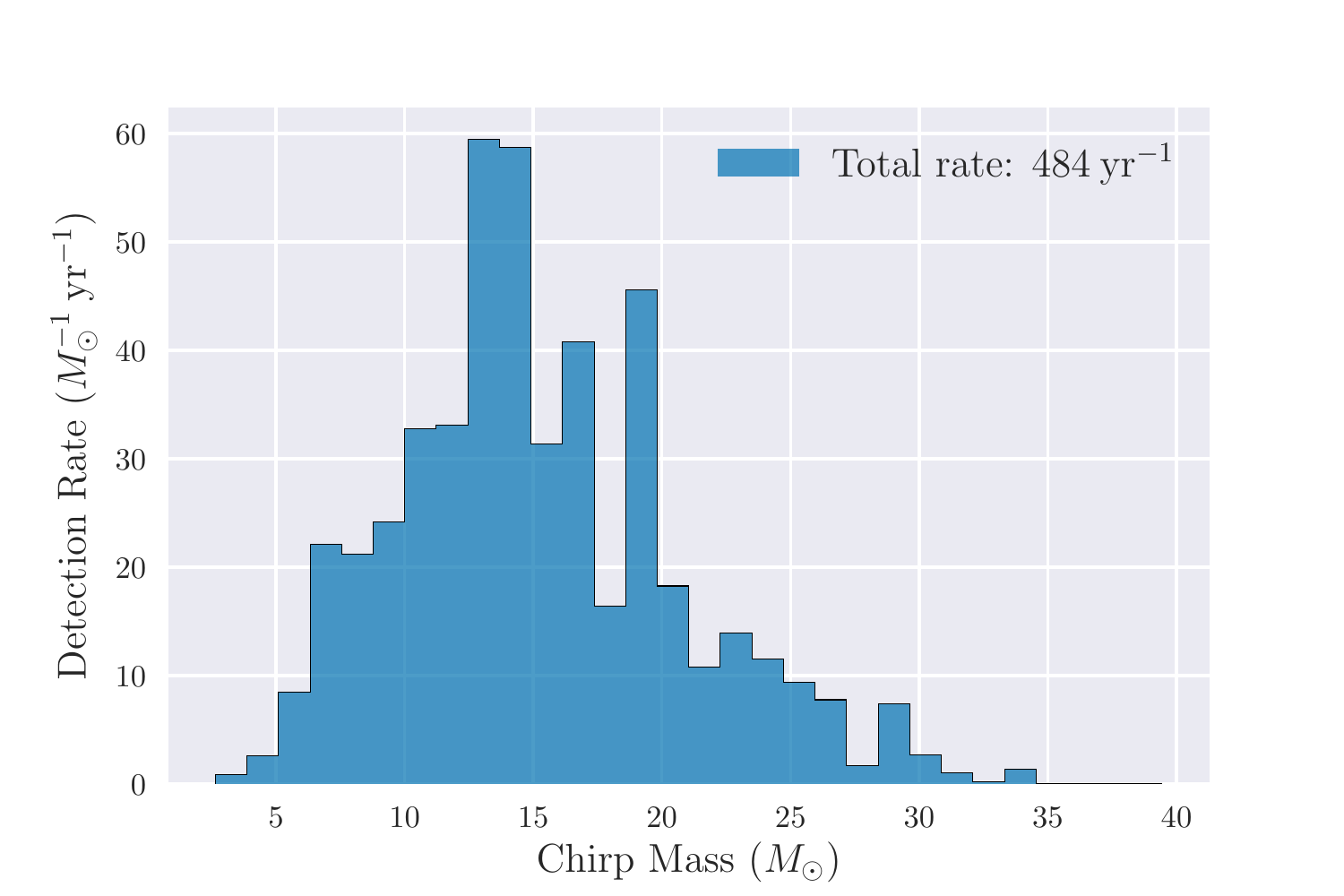}
 \caption{The rate and chirp-mass distribution of the binary black hole coalescences we expect aLIGO to observe at design sensitivity, taking into account cosmic history and selection effects, for the \texttt{COMPAS} fiducial model as described in \citet{stevenson2017formation}. The detection rate is per unit observing time.}
   \label{fig::rates}
\end{center}
\end{figure}

Figure~\ref{fig::rates} shows the rate and chirp-mass distribution of binary black hole mergers detected at aLIGO design sensitivity. The mass distribution is shifted to higher masses relative to the intrinsic merger rate plotted in figure \ref{fig::cosmicRates} because selection effects favour heavier systems which emit louder gravitational-wave signals. Some of the sharp features in this plot are the consequence of simulating systems on a discrete grid of metallicities \citep[cf.][]{dominik2013double}. LBV winds tend to reduce high mass stars to a narrow, metallicity-dependent range of black hole masses.  We discuss the impact of these features in section~\ref{sec::conclusion}. 

\section{The covariance matrix for population parameters}\label{sec::methods}

\subsection{The Fisher information matrix}

The Fisher matrix quantifies the amount of information that a set of observable random variables (in our case, the merger rate and chirp-mass distributions) carries about the parameters (in our case, the four tunable parameters described in section \ref{sec::binaryPopSynth}) of a distribution that models these observables.  

Specifically, the Fisher matrix $F$ for a set of random variables $\mathcal{D}$ (the data) which are dependent on a set of parameters $\{\lambda\}$ is defined element-wise as
\begin{equation}\label{eq::fisherDefinition}
F_{ij} = - \left<\frac{\partial^2 \log\left[\mathcal{L}\left(\mathcal{D}|\{\lambda\}\right)\right]}{\partial \lambda_i \; \partial \lambda_j}\right>,
\end{equation}
where $\mathcal{L}$ is the likelihood function, defined as the probability of acquiring the observed data $\mathcal{D}$ given the model parameters, and the angle brackets indicate an expectation over the data realisation. We introduce the likelihood for our problem in the section below.

Under certain conditions, the inverse of the Fisher matrix gives a lower bound (the Cr\'amer--Rao bound) on the covariance matrix for those dependent parameters \citep{2008PhRvD..77d2001V}; we discuss the regime of validity of the Fisher matrix inverse as an approximation to the covariance matrix in section~\ref{ssec::numberOfObservations}. The covariance matrix tells us how sensitive our data is to a change in the model parameters.
We can also examine which combinations of dependent parameters are degenerate and which combinations yield the greatest information gain.\footnote{This is analogous to identifying the chirp mass as being the best measured combination of masses from gravitational-wave observations.}

The Fisher matrix quantifies the sensitivity of predicted observations to model parameters, and provides a bound on the accuracy of parameter inference.  This approach assumes that the model is correct. The correctness of the model can be evaluated through other means.  For example, model selection can be used to compare distinct models, whether these are different formation channels or different prescriptions for describing the physical processes of binary evolution \citep[e.g.,][]{Mandel:2009nx,2017CQGra..34cLT01V,2017MNRAS.471.2801S,2017ApJ...846...82Z,2017PhRvD..96b3012T}, or model-independent clustering can be used without reference to particular models \citep[e.g.,][]{Mandel:2015,Mandel:2016cluster}.

\subsection{The \texttt{COMPAS} likelihood function}\label{ssec::likelihoodFunction}

For this study we assume that we have a gravitational-wave catalogue of merging binary black holes, formed via the isolated binary evolution channel, and we focus on two observable characteristics of such a dataset: the rate of detections and the distribution of chirp masses for the observed systems.

The likelihood function contains a term for each observational characteristic:
\begin{equation}\label{eq::compasLikelihood}
\log{\mathcal{L}\left( \mathcal{D} |\{\lambda\}\right)} = \log{\mathcal{L}\left(N_{\mathrm{obs}}|\{\lambda\},t_{\mathrm{obs}}\right)} + \log{\mathcal{L}\left(\{ \mathcal{M} \}|\{\lambda\}\right)}.
\end{equation}
The first term is the likelihood of observing binary black holes at a given rate. We assume that observable binary black holes coalesce in the Universe as a Poisson process with rate parameter $\mu$, which is predicted by our population synthesis model, and total number of observations $N_{\mathrm{obs}}$, accumulated in a time $t_{\mathrm{obs}}$. The Poisson likelihood is
\begin{equation}\label{eq::ratesLogLikelihood}
\log{\mathcal{L}\left(N_{\mathrm{obs}}|\{\lambda\},t_{\mathrm{obs}}\right)} = N_{\mathrm{obs}}\log(\mu t_{\mathrm{obs}}) - \mu t_{\mathrm{obs}} - \log(N_{\mathrm{obs}}!).
\end{equation}
The second term is the likelihood of observing a given chirp-mass distribution. As described in section~\ref{sec::modelpredictions}, we have chosen to represent our chirp-mass distribution in bins. In this case the correct likelihood is a multinomial distribution \citep{Stevenson:2015bqa}
\begin{equation}
\log{\mathcal{L}\left(\{ \mathcal{M} \}|\{\lambda\}\right)} = \log(N_{\mathrm{obs}}!) + \sum_k^{K} \left[c_k \log(p_k) - \log(c_k!) \right],
\end{equation}
where $K$ is the number of chirp-mass bins, $c_k$ is the number of observed systems falling into the $k$-th bin with $\sum_k c_k = N_\mathrm{obs}$, and $p_k$ is the probability predicted by the model that a system falls into the $k$-th bin. Thus, $\mu$ and $p_k$ are functions of the tunable model parameters $\lambda$, while $c_k$ and $N_\mathrm{obs}$ are observables. Given the likelihood, we can now calculate the Fisher matrix.

\subsection{Computing the Fisher matrix}\label{ssec::computingFisher}  

In order to compute the Fisher matrix, we need to find the second derivatives of the likelihood with respect to the population parameters and average over the possible observations drawn according to the same likelihood distribution. First differentiating the total-rate log likelihood,
\begin{align}
\frac{\partial^2 \log{\mathcal{L}\left(N_\mathrm{obs}|\{\lambda\}\right)}}{\partial \lambda_i\;\partial\lambda_j} = & \frac{\partial}{\partial \lambda_j}\left[\left(\frac{N_\mathrm{obs}}{\mu}-t_\mathrm{obs}\right)\frac{\partial\mu}{\partial\lambda_i}\right] \nonumber \\
= & -\frac{N_\mathrm{obs}}{\mu^2}\frac{\partial\mu}{\partial\lambda_i}\frac{\partial\mu}{\partial\lambda_j} \nonumber \\
 & + \left(\frac{N_\mathrm{obs}}{\mu}-t_\mathrm{obs}\right)\frac{\partial^2\mu}{\partial\lambda_i\;\partial\lambda_j}.
\end{align}
Meanwhile, differentiating the chirp-mass portion of the log likelihood yields
\begin{align}
\frac{\partial^2 \log{\mathcal{L}\left(\{ \mathcal{M} \}|\{\lambda\}\right)}}{\partial \lambda_i\;\partial\lambda_j} = & \frac{\partial}{\partial \lambda_j}\left(\sum_k^K \frac{c_k}{p_k}\frac{\partial p_k}{\partial \lambda_i}\right) \nonumber \\
= & \sum_k^K \left(-\frac{c_k}{p_k^2}\frac{\partial p_k}{\partial \lambda_i}\frac{\partial p_k} {\partial \lambda_j}+\frac{c_k}{p_k}\frac{\partial^2 p_k}{\partial \lambda_i\; \partial \lambda_j}\right).
\label{eq::rate-full-L}
\end{align}

The expectation value of $N_\mathrm{obs}$ over this Poisson likelihood with rate parameter $\mu t_\mathrm{obs}$ is just $\langle N_\mathrm{obs}\rangle=\mu t_\mathrm{obs}$; similarly, $\langle c_k \rangle = \mu t_\mathrm{obs} p_k$. Therefore, the Fisher matrix is
\begin{equation}
\label{eq::fisherMatrixElements}
F_{ij} =  \mu t_{\mathrm{obs}}\left[\frac{1}{\mu^2}\frac{\partial\mu}{\partial\lambda_i}\frac{\partial\mu}{\partial\lambda_j} + \sum_k^K \frac{1}{p_k}\frac{\partial p_k}{\partial \lambda_i}\frac{\partial p_k}{\partial \lambda_j} \right],
\end{equation}
where we used $\sum_k p_k =1$ to eliminate the second term from Eq.~\eqref{eq::rate-full-L}.  Crucially, this expression contains only first-order derivatives of the observables with respect to the population parameters. These derivatives can be readily and reliably estimated using population synthesis models, as described below.  

\subsection{Evaluating the first derivatives}
\label{ssec::evaluatingDerivatives}

We have shown in Eq.~\eqref{eq::fisherMatrixElements} that the Fisher matrix can be computed using just the first derivatives of the binned rates with respect to the population parameters. To compute derivatives, we simulated binary populations using a suite of variations to the population parameters discussed in section~\ref{ssec:populationParameters}. We used the same set of random seeds to the random number generator in \texttt{COMPAS}, so that for each variation the initial conditions (i.e.\ masses and separation) and random effects (i.e.\ kick directions) remain fixed. This allows us to accurately measure the derivatives by estimating the differential rather than absolute rates, reducing the uncertainty associated with a limited number of simulations.

We made six perturbations to the fiducial model for each population parameter (three negative and three positive). The perturbations were chosen to be sufficiently small that we could reliably estimate first derivatives numerically. A full list of the variations we used can be found in table~\ref{tab::simulationDetails}. For each of the quantities we are differentiating, we have a set of overconstrained simultaneous equations for the first and second derivatives according to the leading terms in the Taylor series, which we can write in matrix form
\begin{equation}\label{eq::taylorSeriesMatrix}
\left(
\begin{array}{c}
\displaystyle
f(\lambda + \Delta_1) - f(\lambda) \\
\vdots \\
f(\lambda + \Delta_6) - f(\lambda)
\end{array}\right)
 = \left(\begin{array}{cc}
\Delta_1 & \displaystyle \frac{1}{2}\Delta_1^2 \\
\vdots & \vdots \\
\Delta_6 & \displaystyle \frac{1}{2}\Delta_6^2 \\
\end{array}\right)
\left(\begin{array}{c}
\displaystyle \frac{\partial f(\lambda)}{\partial \lambda} \\
\; \\
\displaystyle \frac{\partial^2 f(\lambda)}{\partial \lambda^2}
\end{array}\right).
\end{equation}
If we label the three terms in Eq.~\eqref{eq::taylorSeriesMatrix} as $\mathbf{y}$, $\mathbf{X}$ and $\mathbf{\beta}$ respectively, then the maximum-likelihood solution for the derivatives $\mathbf{\hat{\beta}}$ can be computed directly as \citep[][section 9.3]{Anton2000}
\begin{equation}\label{eq::normalEquation}
\mathbf{\hat{\beta}} = (\mathbf{X}^T\mathbf{X})^{-1} \mathbf{X}^T \mathbf{y}.
\end{equation}
We use this approach to compute all of the derivatives in Eq.~\eqref{eq::fisherMatrixElements} and combine them into an estimate of the Fisher matrix. The Fisher matrix can then be inverted to provide the Cr\'amer--Rao lower bound on the covariance matrix of the astrophysical parameters evaluated at the \texttt{COMPAS} fiducial model.

\begin{table}
\centering
  \begin{tabular}{@{} D{.}{.}{3.4} D{.}{.}{1.3} D{.}{.}{1.3} D{.}{.}{1.3} @{}}
  \hline
  \multicolumn{1}{c}{$\sigma_\mathrm{kick}$ [$\mathrm{km\,s}^{-1}$]} & \multicolumn{1}{c}{$\alpha_{\mathrm{CE}}$} & \multicolumn{1}{c}{$f_{\mathrm{WR}}$} & \multicolumn{1}{c}{$f_{\mathrm{LBV}}$} \\
  \hline
  250.0 & 1.00 & 1.00 & 1.50 \\
  \hdashline
  240.0 & 1.00 & 1.00 & 1.50 \\
  244.0 & 1.00 & 1.00 & 1.50 \\
  247.0 & 1.00 & 1.00 & 1.50 \\
  253.0 & 1.00 & 1.00 & 1.50 \\
  256.0 & 1.00 & 1.00 & 1.50 \\
  260.0 & 1.00 & 1.00 & 1.50 \\
  \hdashline
  250.0 & 0.95 & 1.00 & 1.50 \\
  250.0 & 0.97 & 1.00 & 1.50 \\
  250.0 & 0.99 & 1.00 & 1.50 \\
  250.0 & 1.01 & 1.00 & 1.50 \\
  250.0 & 1.03 & 1.00 & 1.50 \\
  250.0 & 1.05 & 1.00 & 1.50 \\
  \hdashline
  250.0 & 1.00 & 0.90 & 1.50 \\
  250.0 & 1.00 & 0.94 & 1.50 \\
  250.0 & 1.00 & 0.97 & 1.50 \\
  250.0 & 1.00 & 1.03 & 1.50 \\
  250.0 & 1.00 & 1.06 & 1.50 \\
  250.0 & 1.00 & 1.10 & 1.50 \\
  \hdashline
  250.0 & 1.00 & 1.00 & 1.45 \\
  250.0 & 1.00 & 1.00 & 1.47 \\
  250.0 & 1.00 & 1.00 & 1.49 \\
  250.0 & 1.00 & 1.00 & 1.51 \\
  250.0 & 1.00 & 1.00 & 1.53 \\
  250.0 & 1.00 & 1.00 & 1.55 \\
  \hline
  \end{tabular}
 \caption{The $25$ population-parameter variations used in this paper. The population parameters are described in section~\ref{ssec:populationParameters}: $\sigma_\mathrm{kick}$ is the dispersion parameter for a Maxwellian used to draw the magniutde of natal kicks from Eq.~\eqref{eq:kick_maxwell}; $\alpha_\mathrm{CE}$ is the efficiency of common-envelope ejection from Eq.~\eqref{eq:alpha_ce}; $f_\mathrm{WR}$ is the multiplier for Wolf--Rayet wind mass loss from Eq.~\eqref{eq:mdot_wr}, and $f_\mathrm{LBV}$ is the multiplier for luminous blue variable mass loss described in Eq.~\eqref{eq:mdot_lbv}. Our fiducial model appears in the top row. For each of these population parameter combinations we also varied metallicity. We used $12$ different metallicities, which were evenly spaced in the log between $0.005 Z_{\odot}$ and $Z_{\odot}$, where we use a solar metallicity $Z_{\odot} = 0.02$. We therefore had a total of $300$ model variations. We simulated $1,197,989$ binaries for each of these variations.}
   \label{tab::simulationDetails}
\end{table}

\subsection{Measurement uncertainty}\label{ssec::measurementUncertainty}

The measurements of chirp masses will be subject to a certain amount of measurement uncertainty. We use a simplified treatment of this measurement uncertainty based on the methodology of \citet{Gair2010LisaEmri}, see their appendix A. We assume that the probability of finding a system in an incorrect bin is given by a Gaussian distribution about the centre of the correct bin into which the system was placed in the simulation.

\begin{figure}
\begin{center}
 \includegraphics[width=0.5\textwidth]{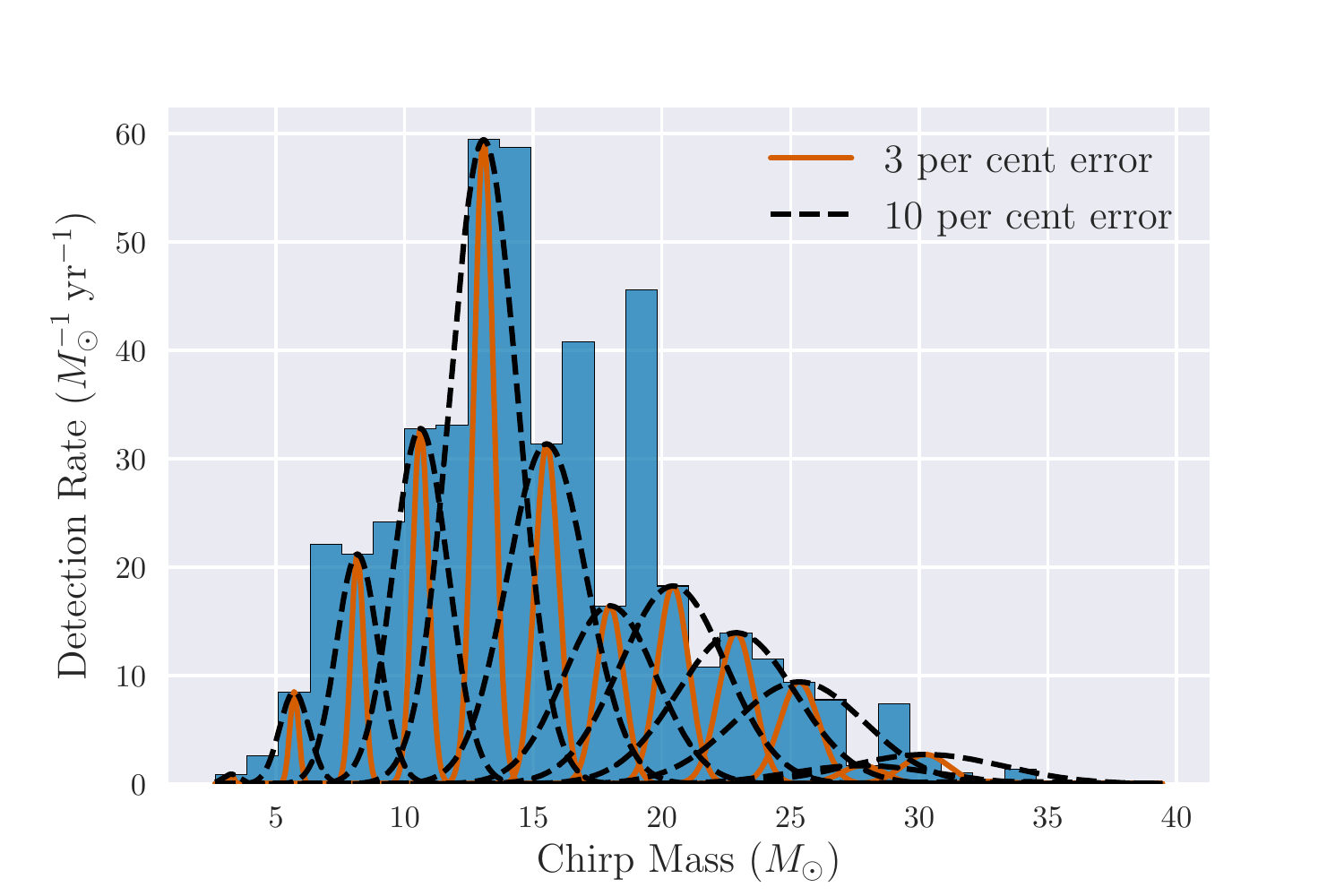}
 \caption{An illustration of how we include measurement errors in our analysis. A Gaussian is centred on each bin, with a standard deviation proportional to the value at the centre of that bin. That bin's counts are then distributed to other bins according to the fraction of that Gaussian falling in each bin.}
   \label{fig::measurementErrorIllustration}
\end{center}
\end{figure}

Let $f_i$ be the fraction of system predicted by the simulation to lie in the  $i$-th bin, which is centred on chirp mass $\mu_i$ and has left and right edges at chirp masses  $\mu_i^-$ and  $\mu_i^+$, respectively. Then the probability $p_i$ of observing a system in the $i$-th bin is
\begin{equation}\label{eq::measurementErrors}
p_{i} = \sum_{j=1}^{K} \frac{f_{j}}{\sqrt{2\pi \sigma_j^2}} \int_{\mu_i^-}^{\mu_i^+} \dd x\; \exp\left[\frac{-(x-\mu_j)^2}{2\sigma_j^2} \right],
\end{equation}
\noindent
where $\sigma_i$ is the standard deviation of the measurement in the $i$-th bin. In the limit of $\sigma_i$ tending to zero, we recover perfect measurement accuracy, $p_i=f_i$. An illustration of this treatment of the measurement errors is presented in figure~\ref{fig::measurementErrorIllustration}.

The chirp-mass measurement uncertainty depends strongly on the total mass of the source, with the most massive sources spending the fewest inspiral cycles in band, leading to the largest measurement uncertainty \citep[e.g.,][]{BBH:O1}.  It also scales inversely with the source SNR. Here, we crudely approximate this as a fixed fractional uncertainty on the chirp mass of $3~\mathrm{per\ cent}$ \citep[cf.,][]{Mandel:2016cluster,2017PhRvD..95f4053V}. We therefore modify the binned rates according to Eq.~\eqref{eq::measurementErrors}, using a standard deviation $\sigma_i = 0.03\mu_i$.

This method of incorporating measurement errors is a simplification. The formally correct approach would be to incorporate them on a per-system basis, which would involve a modification of the likelihood function. Performing the analysis in this way would correctly account for correlations between bins, whereas in the simplified approach bins are modified independently, losing information and slightly swelling the uncertainty.

\subsection{Uncertainty quantification}\label{ssec::uncertaintyQuantification}

The rate derivatives used to compute the Fisher matrix at the \texttt{COMPAS} fiducial model depend on the particular population realisation used in the calculation.  We quantify the impact of simulation realisation noise, due to the finite number of simulated binaries, with bootstrapping.  We recompute the Fisher matrix by re-creating data sets of the same size as the original simulated data set by drawing samples from it with replacement.

By repeating this process many times and observing the spread in the results, we can observe how much the point estimates change under different population realisations (different sets of binary initial conditions). 
Our full dataset consists of $359,396,700$ binary simulations, which consists of the same set of $1,197,989$ ZAMS binaries evolved under each of $300$ different model variations (the $25$ population parameter combinations listed in table~\ref{tab::simulationDetails}, each simulated at the $12$ different metallicities shown in figure~\ref{fig::MSSFR}). To generate one bootstrap sample Fisher matrix:
\begin{enumerate}
	\item We randomly choose $1,197,989$ initial conditions, with replacement, from our original set of initial conditions. 
	\item For each of the $25$ population parameter combinations in table~\ref{tab::simulationDetails}, we find the systems from the bootstrap initial conditions which become merging binary black holes, and calculate their total rate and chirp-mass distribution (taking into account cosmic history, selection effects and measurement uncertainty).
	\item We use Eq.~\eqref{eq::taylorSeriesMatrix} and Eq.~\eqref{eq::normalEquation} to compute the derivatives of the total rate and chirp-mass distribution bin heights, with respect to each population parameter.
	\item We use these derivatives to compute the Fisher matrix, using Eq.~\eqref{eq::fisherMatrixElements}.
\end{enumerate}
We repeat the above steps $1500$ times in order to express the uncertainty coming from the realisation of the initial conditions, i.e.\ from the simulation statistical fluctuations. 
In principle, this model uncertainty could be overcome with more simulations, unlike the more fundamental uncertainties stemming from a finite number of observations and chirp-mass measurement uncertainty. We discuss the relative contributions of these sources of uncertainty in section~\ref{sec::conclusion}.

\section{Results and discussion}\label{sec::results}

Using the method described in section~\ref{sec::methods} we computed the elements of the Cr\'amer-Rao lower bound on the covariance matrix for the population parameters $\sigma_{\mathrm{kick}}$, $\alpha_{\mathrm{CE}}$, $f_{\mathrm{LBV}}$ and $f_{\mathrm{WR}}$. We computed simulation uncertainties on these elements by taking $1500$ bootstrap samples from the $1,197,989$ sets of initial conditions simulated for the binaries, specifically varying the metallicities, initial masses and separations. Using these results we are able to explore what can be learned about these population parameters using gravitational-wave observations of binary black holes. Results are presented for $N_{\mathrm{obs}}=1000$ observations, a sufficiently large number to ensure the validity of our results; we discuss the effect of changing the number of observations in section~\ref{ssec::numberOfObservations}.

\begin{figure}
\begin{center}
 \includegraphics[width=0.5\textwidth]{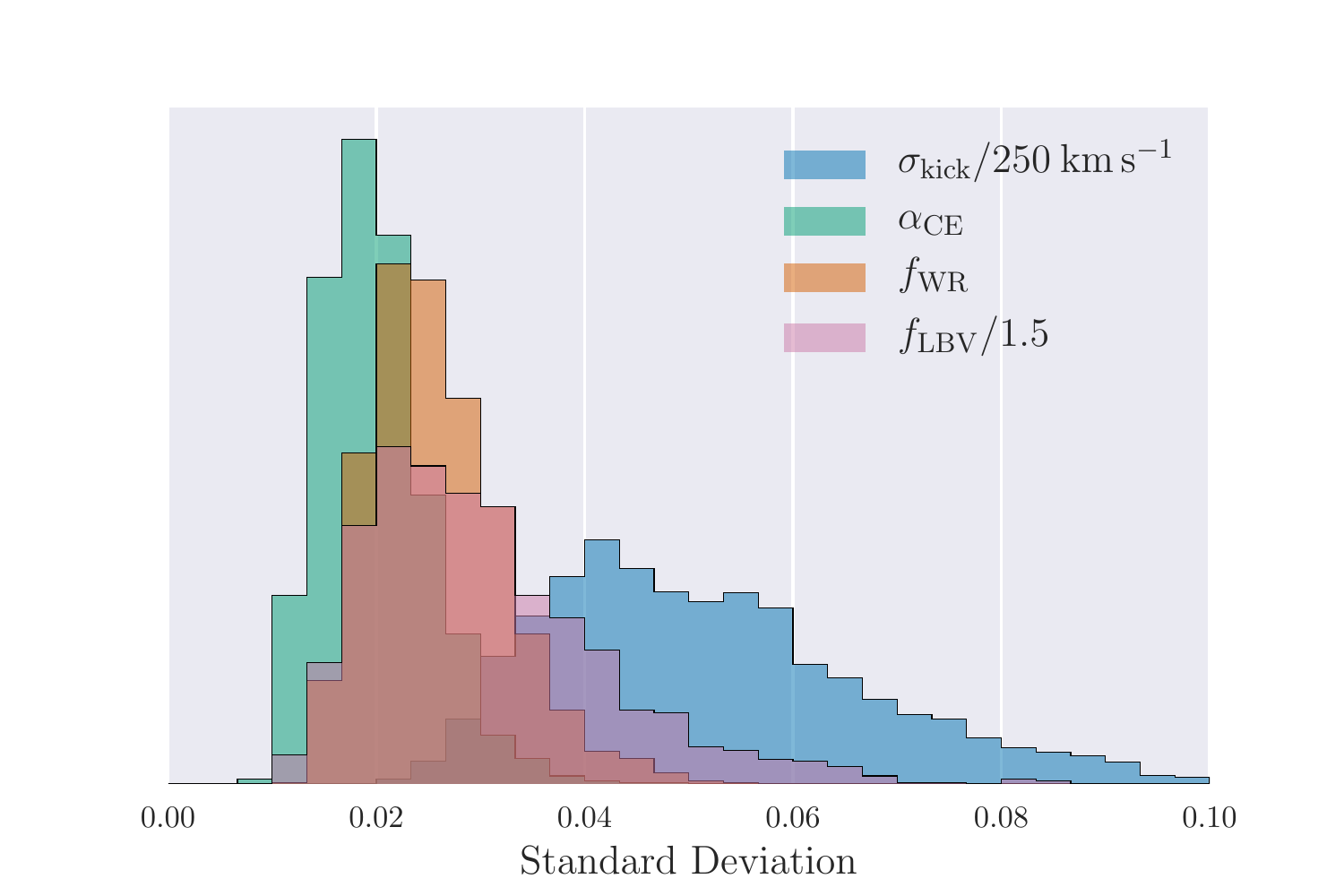}
 \caption{The inferred measurement accuracy for each of the four population parameters after observing $1000$ systems, as estimated by taking the square root of the diagonal elements of the estimated covariance matrices for each of the $1500$ bootstrapped sets. The histograms are normalised such that they all have the same area.}
   \label{fig::standardDeviations}
\end{center}
\end{figure}

Figure~\ref{fig::standardDeviations} shows the distribution of standard deviations of each of the population parameters. We see that it will be possible to measure $\alpha_{\mathrm{CE}}$, $f_{\mathrm{LBV}}$ and $f_{\mathrm{WR}}$ with fractional accuracies of $\sim 2~\mathrm{per\ cent}$ after $1000$ observations. We will be less sensitive to the value of $\sigma_{\mathrm{kick}}$. This is an expected result, since the natal kicks of black holes are reduced according to Eq.~\eqref{eq::kickFallback}, and many of the more massive ones do not get a kick at all. 

The fractional uncertainties on all of the parameters are quantities of order $N_\mathrm{obs}^{-1/2} \approx 0.03$ for $N_\mathrm{obs}=1000$. Varying the parameters by their full dynamic range would change the rate by $\mathcal{O}(N_\mathrm{obs})$.  For example, reducing $\alpha_\mathrm{CE}$ from $1$ to $0$ would make binary black hole formation through a common-envelope phase impossible, reducing the expected number of detections from $N_\mathrm{obs}$ to $\sim0$. 

The measurement accuracy with which the tunable population parameters can be inferred using $1000$ gravitational-wave observations can be alternatively interpreted from the perspective of model selection.For example, the median of the distribution for the standard deviation of $\alpha_{\mathrm{CE}}$ is $\sim 0.02$. Therefore, if $\alpha_{\mathrm{CE}}$ different from the fiducial value by $6~\mathrm{per\ cent}$, the fiducial model could be ruled out with a confidence of $\sim 3\sigma\approx 99.7~\mathrm{per\ cent}$.

\begin{figure}
\begin{center}
 \includegraphics[width=0.5\textwidth]{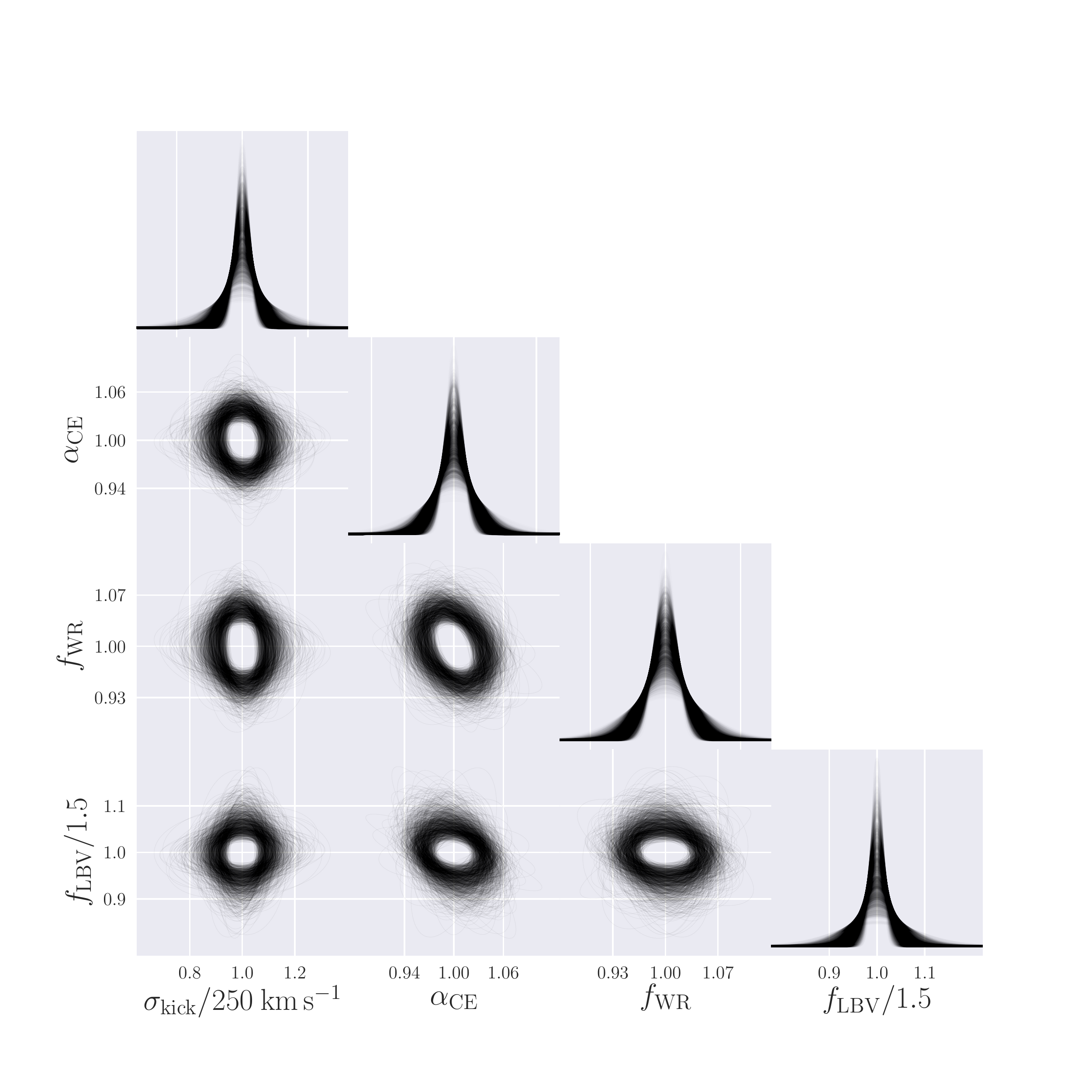}
 \caption{$1500$ bootstrap samples of the marginalised univariate distributions and bivariate $90~\mathrm{per\ cent}$ confidence intervals from the Cr\'amer--Rao lower bound on the covariance matrix for the \texttt{COMPAS} population parameters. The univariate distributions are the Gaussian distributions corresponding to the standard deviations of figure~\ref{fig::standardDeviations}, and have been normalised to have the same area.}
   \label{fig::covarianceCorner}
\end{center}
\end{figure}

\begin{figure}
\begin{center}
 \includegraphics[width=0.5\textwidth]{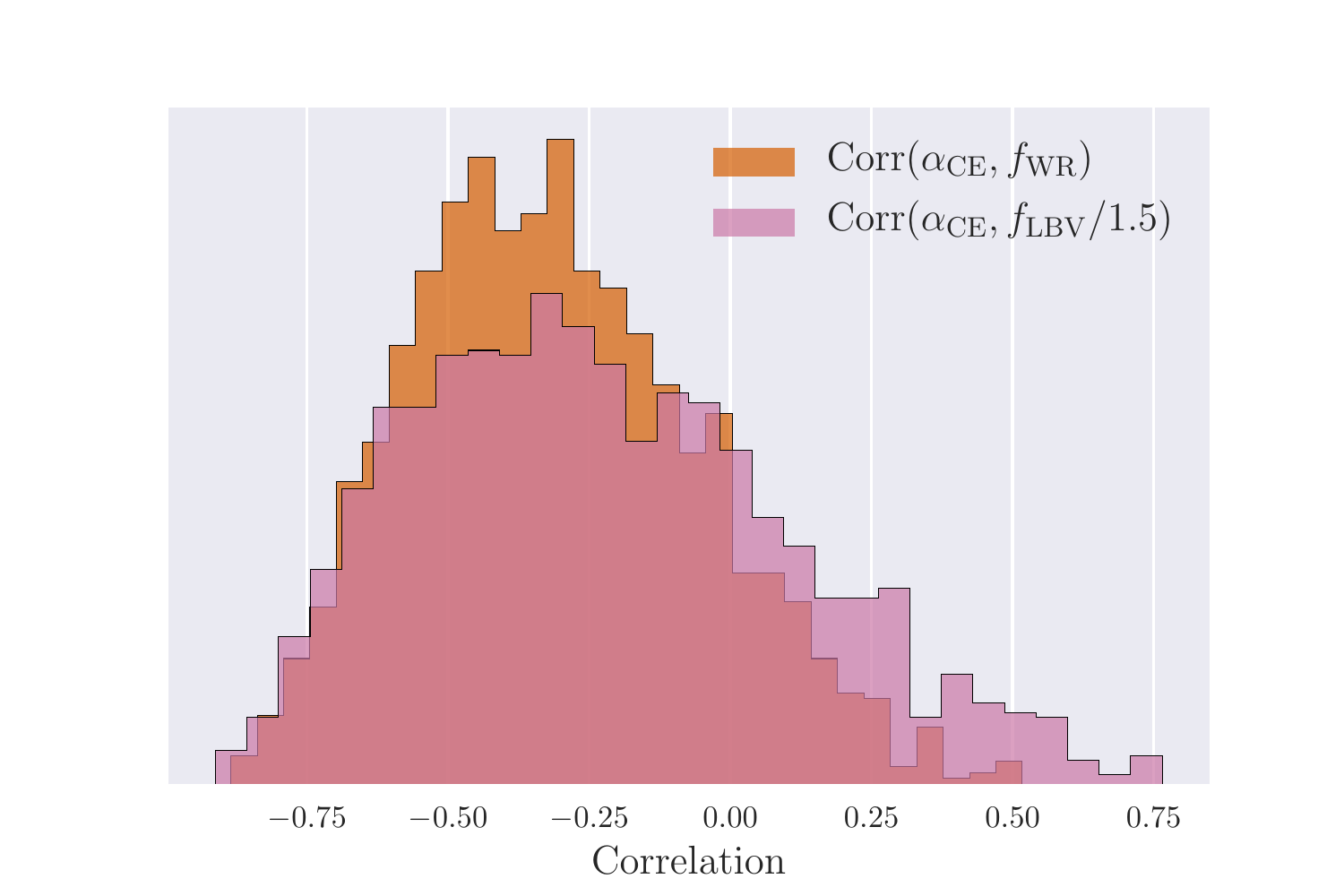}
 \caption{Distribution of correlations between $\alpha_{\mathrm{CE}}$ and each of $f_{\mathrm{LBV}}$ and $f_{\mathrm{WR}}$. The histograms have been normalised to have the same area.}
   \label{fig::correlations}
\end{center}
\end{figure}

We can examine the full multivariate normal behaviour of the population parameters. Figure~\ref{fig::covarianceCorner} shows marginalised univariate distributions and bivariate projections of the $90~\mathrm{per\ cent}$ confidence interval for each of the bootstrap samples. This plot shows that most pairwise correlations between most population parameters are negligible. Figure~\ref{fig::correlations} shows the correlations between $\alpha_{\mathrm{CE}}$ and $f_{\mathrm{WR}}$, and between $\alpha_{\mathrm{CE}}$ and $f_{\mathrm{LBV}}$. Bootstrapping indicates an $88~\mathrm{per\ cent}$ confidence that $\alpha_{\mathrm{CE}}$ and $f_{\mathrm{WR}}$ are anti-correlated. Increasing $\alpha_{\mathrm{CE}}$ increases the efficiency with which orbital energy is transferred into the common envelope. An increased efficiency means that there will be less tightening of the binary, so fewer systems will come sufficiently close together to merge within a Hubble time. Losing mass through winds widens the orbit, meaning that increasing the Wolf--Rayet wind mass-loss rate creates more systems which are too wide to merge within a Hubble time. Increased mass loss also results in the black holes being less massive, therefore increasing the time required for them to merge through gravitational-wave emission from a given initial separation \citep{Peters:1964}. These correlations mean that increasing (or decreasing) both $\alpha_{\mathrm{CE}}$ and $f_{\mathrm{WR}}$ would compound the effect on the rates, so their bivariate distribution (in figure~\ref{fig::covarianceCorner}) is narrower in this direction. Conversely, the effects of increasing one whilst decreasing the other would partially cancel out, and thus the bivariate distribution is wider in that direction. The confidence in the anti-correlation between $\alpha_{\mathrm{CE}}$ and $f_{\mathrm{LBV}}$ is only $76~\mathrm{per\ cent}$, and there is insufficient evidence for correlation between other parameter pairs.

\subsection{Information from the total detection rate}

To gain further insight into the correlations between the inferred parameters, 
we now consider what we could learn about the population parameters by considering only the total rate at which gravitational waves are observed. It is impossible to constrain the four-dimensional population parameter vector considered in this paper with a single observable, the binary black hole detection rate. In this case, all that can be learned about the population parameters is the value of some linear combination of them. 

We construct a detection rate Fisher matrix, using only the total rate log likelihood of Eq.~\eqref{eq::ratesLogLikelihood},
\begin{equation}
F^{\mathrm{RO}}_{ij} = \frac{t_{\mathrm{obs}}}{\mu}\frac{\partial\mu}{\partial\lambda_i}\frac{\partial\mu}{\partial\lambda_j},
\end{equation}
\noindent
and perform an eigendecomposition. We expect to see that there is only one eigenvector whose eigenvalue is non-zero. We verified that this is true for all $1500$ of our bootstrap samples, which provided a useful sanity check of our results. 

\begin{figure}
\begin{center}
 \includegraphics[width=0.5\textwidth]{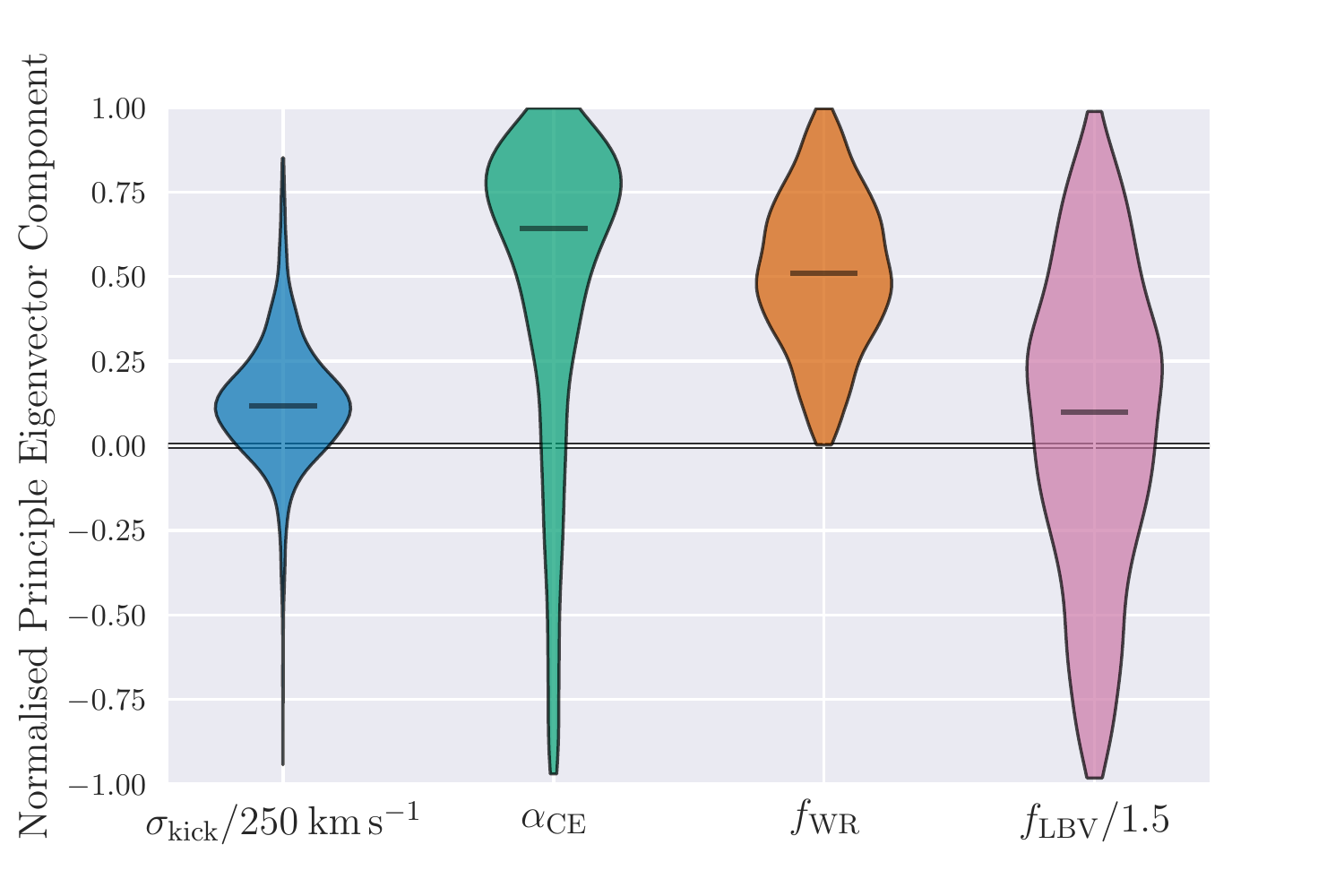}
 \caption{Violin plot showing components of the normalised principal eigenvector of the Fisher matrices calculated using only the total detection rate. The coloured regions give the bootstrapped distribution of the principle eigenvector direction, with medians marked in black.}
   \label{fig::ratesOnlyDirection}
\end{center}
\end{figure}

Next, by examining the eigenvector whose eigenvalue is non-zero, we can find the linear combination of population parameters to which we are sensitive. Figure~\ref{fig::ratesOnlyDirection} shows a univariate representation of this direction (with its distribution from bootstrapping over simulations). The components of the vector parallel to $f_{\mathrm{LBV}}$ and $\sigma_{\mathrm{kick}}$ axes are broadly consistent with zero.  Most of the information learned solely from the total detection rate is in the $\alpha_{\mathrm{CE}}$--$f_{\mathrm{WR}}$ plane. The fact that both values are simultaneously positive implies that they are correlated; this is the same correlation as was discussed at the beginning of this section.

\begin{figure}
\begin{center}
 \includegraphics[width=0.5\textwidth]{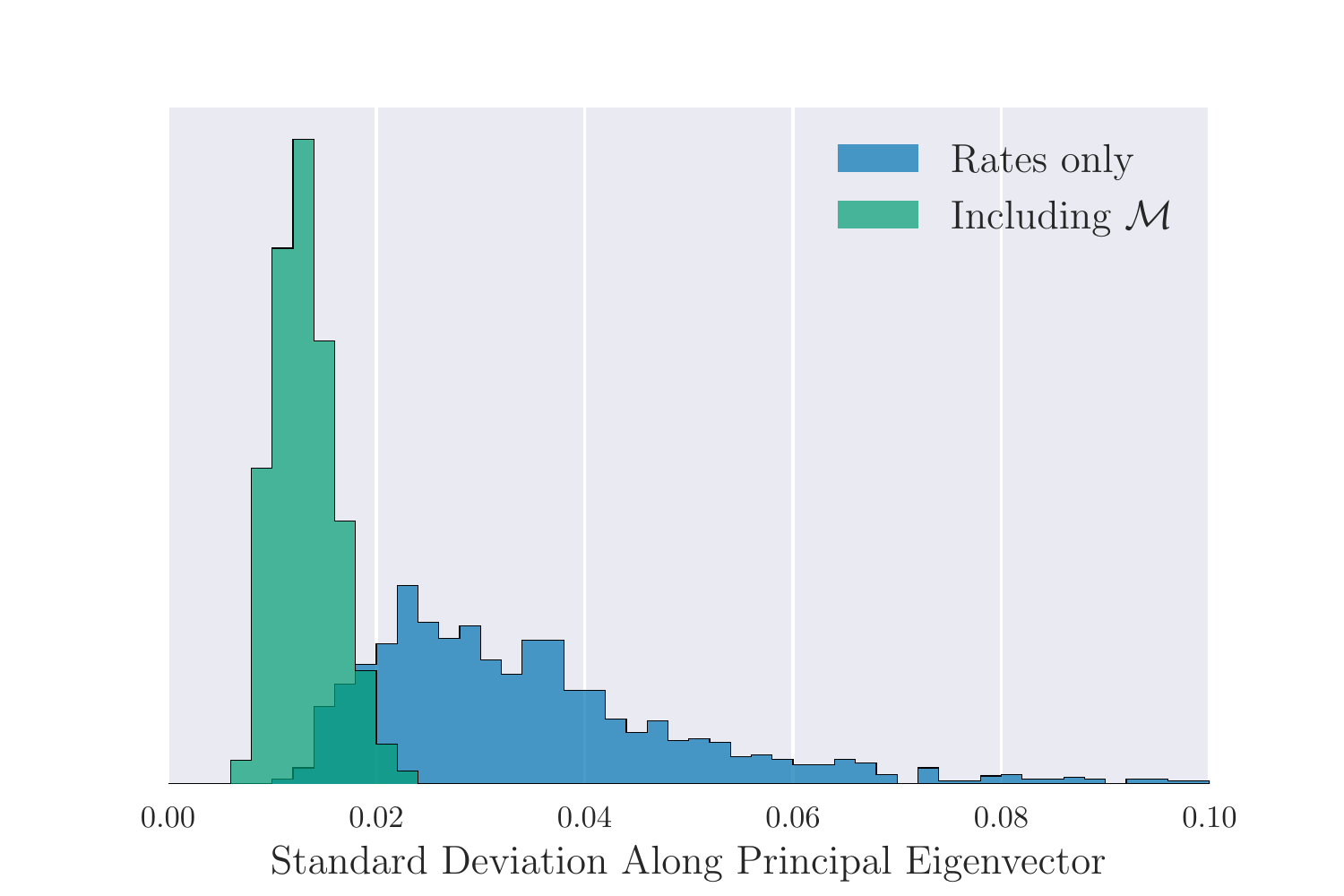}
 \caption{Distribution of the standard deviation of the particular linear combination of population parameters corresponding to the principal eigenvector of the total detection rate Fisher matrix.  The measurement accuracy is computed using only information from the total rate (blue) and after including information from the chirp-mass distribution (green). The distributions come from considering all 1500 bootstrapped sets.}
   \label{fig::ratesOnlyStandardDeviation}
\end{center}
\end{figure}

Whilst we can only measure this specific combination of population parameters using only the total detection rate, we can constrain parameter combinations in the $\sim \alpha_\mathrm{CE}+f_\mathrm{WR}$ direction to within a few per cent from the total rate. Figure~\ref{fig::ratesOnlyStandardDeviation} shows the standard deviation along the line defined by this combination of population parameters $a^{-1/2}$, where $a$ is the principal eigenvalue. This can be interpreted in the same way as the standard deviations in figure~\ref{fig::standardDeviations}, and matches the expected value of $\mathcal{O}(N_\mathrm{obs}^{-1/2})$. We see that if this combination of population parameters differed from our fiducial values by more than a few per cent, we would be able to confidently rule our model out after $1000$ observations. However, we also see from figure~\ref{fig::ratesOnlyStandardDeviation} that including the chirp-mass distribution would significantly improve measurements of this parameter combination.

\subsection{Number of observations}
\label{ssec::numberOfObservations}

The expected number of observations only appears as a multiplicative term in Eq.~\eqref{eq::fisherMatrixElements}, so that the standard deviations in figure~\ref{fig::standardDeviations} simply scale as $N_{\mathrm{obs}}^{-1/2}$. However, the results presented here are predicated on the assumption that the inverse of the Fisher information matrix is a good approximation to the covariance, and not just a lower bound. This in turn requires the likelihood to be approximately Gaussian, i.e.\ the linear single approximation \citep[LSA;][]{2008PhRvD..77d2001V} should hold. 
Only if the predicted parameter uncertainties are smaller than the neighbourhood in which the LSA is valid does the Fisher matrix provide a self-consistent estimate of the accuracy of parameter inference. This effectively sets a minimal threshold on the number of observations required for self-consistency in our estimates.

When computing the derivatives, as described in section~\ref{ssec::evaluatingDerivatives}, we measure the terms in a Taylor expansion of an observable (binned) rate $f$ as a function of the population parameter $\lambda$,
\begin{equation}
f(\lambda+\Delta ) - f(\lambda) \approx \Delta  f'(\lambda) +\frac{\Delta^2 }{2}f''(\lambda).
\end{equation}
In order to verify the validity of the LSA, we need to check that each $f$ is indeed linear when $\Delta$ is of the same order as the computed standard deviations for the population parameters. We require that the linear term is dominant in the Taylor series, so that
\begin{equation}
f'(\lambda) \gg \frac{\Delta}{2}f''(\lambda).
\end{equation}

We find $N_\mathrm{obs}=1000$ to be a sufficient lower limit on the number of observations necessary to ensure the LSA is valid. At $1000$ observations, the best measured combination of parameters is constrained at the per cent level, and this will continue to improve as we expand the catalogue of observations.

For smaller numbers of observations, the LSA will break down. The probability distribution for the model parameters may no longer be a multi-dimensional Gaussian so the Fisher matrix is likely to under-estimate the inference uncertainty.  

\section{Conclusions}
\label{sec::conclusion}

We have, for the first time, quantitatively analysed how accurately gravitational-wave observations of binary black hole mergers will constrain binary population synthesis models described by a multi-dimensional parametrisation. When ground-based detectors have accumulated $1000$ observations of merging binary black holes, we have shown that we will measure binary population synthesis model parameters with an accuracy of a few per cent. Equivalently, we will be able to distinguish models for which the population parameters only differ by a few per cent.

Our analysis accounts for three distinct sources of uncertainty in the inference of population parameters using gravitational-wave observations. The first is due to the finite number of observations. We show when the linear signal approximation holds (section~\ref{ssec::numberOfObservations}), the accuracy with which population parameters can be inferred scales with the inverse square root of the number of observations.  The second is the chirp-mass measurement uncertainty in individual observations. We only model this approximately (section~\ref{ssec::measurementUncertainty}) but find that it is unlikely to be limiting factor in inference. The third source of uncertainty is simulation uncertainty: the accuracy in predicted detection rate and chirp-mass distribution is limited by the finite number of \texttt{COMPAS} simulations. This uncertainty, which we quantify with bootstrapping (section~\ref{ssec::uncertaintyQuantification}), is only limited by computational cost, and be reduced indefinitely with more simulations or more efficient sampling \citep[e.g.,][]{Andrews:2017}.   

There is, of course, potential systematic uncertainty in the binary evolution models themselves: for example, it is probable that the $\alpha_\mathrm{CE}$ parameter is not universal, as assumed here, but depends on the binary properties during the common-envelope phase.  Model inference techniques such as those described here should be combined with model verification and with weakly modelled inference \citep[e.g.,][]{Mandel:2016cluster}.    

We show the expected detection rate and chirp-mass distribution of merging binary black holes in figure~\ref{fig::rates}.  The sharp features in the chirp-mass distribution are due to only simulating systems at a small number ($12$) of metallicities, replacing the integral over metallicity in Eq.~\eqref{eq::cosmicIntegral} with a discrete sum. Mass loss, particularly during the luminous blue variable phase, leads to a pile up of black hole masses from the most massive stars at particular metallicity-dependent values.  The subsequent discrete sum over metallicities overpopulates some bins in the chirp-mass distribution relative to neighbouring bins \citep[cf.][]{dominik2013double}.  This can impact our results, causing us to overstate the accuracy with which we will be able to measure population parameters.  This issue can be addressed in the future by interpolating model predictions over metallicity \citep[e.g., using Gaussian process emulators as described by][]{Barrett:2017tug}, producing a smooth set of predictions.

Our primary intention with this paper was to introduce a methodology for evaluating the accuracy with which astrophysical model parameters can be estimated based on the rates and properties of observed transients. We considered a four-dimensional parameter space, but the number of dimensions is limited only by computational cost.  It is also straightforward to add more terms than just the chirp-mass distribution to Eq.~\eqref{eq::compasLikelihood} in order to investigate other observable characteristics of binary populations such as mass ratios and spins \citep[e.g.,][]{2017MNRAS.471.2801S,2017PhRvD..96b3012T,2017ApJ...846...82Z}.  Furthermore, this analysis can be used for other populations than observations of binary black hole mergers via gravitational-waves in this paper.  Other observed populations, such as Galactic binary pulsars, X-ray binaries, Wolf--Rayet stars, short gamma-ray bursts or luminous red novae \citep[for a review, see][]{DeMarcoIzzard:2017}, can provide orthogonal constraints on the parameters governing binary evolution (cf.\ figure \ref{fig::ratesOnlyStandardDeviation}). Over the coming decade, such measurements will help us to accurately determine the physics of massive binary evolution.

\section*{Acknowledgments}

We thank Floor Broekgaarden and Jonathan Gair for useful comments and discussions, and the anonymous referee for constructive suggestions. We acknowledge Science and Technology Facilities Council (STFC) who supported this work. AVG acknowledges support from Consejo Nacional de Ciencia y Tecnologia (CONACYT). SS acknowledges support from the Australian Research Council Centre of Excellence for Gravitational Wave Discovery (OzGrav), through project number CE170100004.

\bibliographystyle{mnras}
\bibliography{jimBib}

\label{lastpage}

\end{document}